\documentclass[10pt,aps,prd,twocolumn,superscriptaddress,floatfix,amsmath,amssymb,amsfonts,longbibliography,nofootinbib]{revtex4-2}

\usepackage{placeins}

\usepackage{comment}
\usepackage[normalem]{ulem}
\usepackage[english]{babel}
\usepackage{graphicx}% Include figure files
\usepackage{dcolumn}% Align table columns on decimal point
\usepackage{bm}% bold math
\usepackage{blindtext}
\usepackage{verbatim}
\usepackage{mathrsfs}
\usepackage{musicography}
\usepackage{amsmath}
\usepackage{dsfont}
\usepackage{mathrsfs}
\usepackage{amssymb}
\usepackage{amsthm}
\usepackage{cancel}
\usepackage{physics}
\usepackage{epstopdf}
\usepackage{mathtools}
\usepackage{color}
\usepackage{epsfig}
\usepackage{pst-grad} % For gradients
\usepackage{pst-plot} % For axes
\usepackage{hyperref}
\usepackage{verbatim}
\usepackage{afterpage}
\usepackage{sidecap}
\hypersetup{
    colorlinks=true,
    linkcolor=blue,
    filecolor=red,      
    urlcolor=cyan,
}
\usepackage{tikzsymbols}
\newcommand{\mf}{\mathsf}%\mathsf is too long
\newcommand{\modify}[1]{\textcolor{black}{#1}}
\newcommand{\ii}{\mathrm{i}}
\newcommand{\END}{\color{black}}%
\newcommand{\funnyx}{{\scriptstyle{\bm{\mathfrak{X}}}}}
\newcommand{\barsf}[1]{\bar{\mathsf{#1}}}
\newcommand{\timelike}{\musNatural}
\newcommand{\lightlike}{\rotatebox[origin=c]{135}{\musNatural}}
\newcommand{\tinyspace}{\mspace{1mu}}
\newcommand{\bignorm}[1]{\bigl\lvert\!\bigl\lvert\tinyspace #1 \tinyspace\bigr\rvert\!\bigr\rvert}
\newcommand{\spacelike}{\rotatebox[origin=c]{90}{\musNatural}}
\newcommand{\shah}[1]{{\color{orange}\bf [Shahnewaz: #1]}}
\newcommand{\carol}[1]{{\color{blue}\bf [Carol: #1]}}
\newcommand{\edu}[1]{{\color{magenta}\bf [Edu: #1]}}

\renewcommand\qedsymbol{}

\newcommand{\kittyket}
{\ket{\raisebox{-.43ex}{\SchrodingersCat{1}}\!\!}}

\newcommand{\kittybra}
{\bra{\raisebox{-.43ex}{\SchrodingersCat{1}}\!\!}}

\DeclareMathOperator*{\sumint}{%
\mathchoice%
  {\ooalign{$\displaystyle\sum$\cr\hidewidth$\displaystyle\int$\hidewidth\cr}}\textbf{}
  {\ooalign{\raisebox{.14\height}{\scalebox{.7}{$\textstyle\sum$}}\cr\hidewidth$\textstyle\int$\hidewidth\cr}}
  {\ooalign{\raisebox{.2\height}{\scalebox{.6}{$\scriptstyle\sum$}}\cr$\scriptstyle\int$\cr}}
  {\ooalign{\raisebox{.2\height}{\scalebox{.6}{$\scriptstyle\sum$}}\cr$\scriptstyle\int$\cr}}
}

\begin{document}

\title{Decoherence from quantum spacetime noise: An open-systems framework with application to neutrino oscillations}

%\title{Neutrino Decoherence in \texorpdfstring{\(\kappa\)}{kappa}-Minkowski Quantum Spacetime: An Open Quantum Systems Paradigm}
%\title{Decoherence from Quantum Spacetime Fluctuations: An Open System Perspective in \texorpdfstring{\(\kappa\)}{kappa}-Minkowski Geometry}
\author{Partha Nandi}
\email{pnandi@sun.ac.za}
\affiliation{Department of Physics, University of Stellenbosch, Stellenbosch, 7600, South Africa}
\affiliation{National Institute of Theoretical and Computational Sciences (NITheCS), Stellenbosch, 7604, South Africa}

\author{Tiasha Bhattacharyya}
\email{tiashab.phy@gmail.com}
\affiliation{Department of Physics, Diamond Harbour Women’s University, D.H. Road, Sarisha 743368, West Bengal, India}

\author{A. S. Majumdar}
\email{archan@bose.res.in}
\affiliation{S. N. Bose National Centre for Basic Sciences, Salt Lake, Kolkata 700106, India}

\author{Graeme Pleasance}
\email{gpleasance1@gmail.com}
\affiliation{Department of Physics, University of Stellenbosch, Stellenbosch, 7600, South Africa}
\affiliation{National Institute of Theoretical and Computational Sciences (NITheCS), Stellenbosch, 7604, South Africa}

\author{Francesco Petruccione}
\email{petruccione@sun.ac.za}
\affiliation{Department of Physics, University of Stellenbosch, Stellenbosch, 7600, South Africa}
\affiliation{National Institute of Theoretical and Computational Sciences (NITheCS), Stellenbosch, 7604, South Africa}
\affiliation{School of Data Science and Computational Thinking, University of Stellenbosch, Stellenbosch 7600, South Africa}

\begin{abstract}
We present a general open-quantum-systems framework to model decoherence induced by stochastic Planck-scale fluctuations of spacetime, focusing on the $\kappa$-Minkowski noncommutative geometry as a representative quantum-gravity scenario. Treating the deformation parameter as Gaussian white noise, we derive a Lindblad-type master equation applicable to arbitrary quantum systems and obtain a distinctive inverse-energy scaling of the decoherence rate, $\Gamma \propto E^{-4}$. As an illustrative example, we analyze a three-level system motivated by neutrino flavor oscillations and derive closed-form expressions for survival and transition probabilities with spacetime-induced damping. The $E^{-4}$ scaling contrasts sharply with the positive power laws often invoked in quantum-gravity phenomenology and predicts negligible decoherence for high-energy neutrinos—consistent with IceCube observations—while implying that the strongest effects arise in the extreme low-energy regime. In this context, the sub-eV-scale energies characteristic of the cosmic neutrino background provide a natural infrared benchmark for illustrating the enhanced sensitivity to quantum-spacetime fluctuations. Our results establish a unified formalism connecting quantum-information methods, open-system dynamics, and quantum-spacetime phenomenology, thereby offering a framework for exploring potential signatures of Planck-scale physics in future low-energy neutrino studies.

\end{abstract}

\maketitle

\section{Introduction}

Quantum decoherence has been a central topic of investigation for more than five decades, initially introduced by Zeh to explain the quantum-to-classical transition \cite{Zeh:1970zz}. It describes how quantum systems lose their characteristic properties, such as superposition and entanglement, due to interactions with an external environment. This process leads to the emergence of classical behavior and the apparent loss of quantum coherence on macroscopic scales \cite{Zurek:2003zz, Schlosshauer:2019ewh, Schlosshauer:2003zy, Hornberger, Anastopoulos:2000hg,as1,as2,as3}.

Concurrently, modern physics grapples with two unresolved challenges: the unification of gravity \cite{gg} and quantum mechanics \cite{ori6, ParthaNandi:2025kwy}, especially in relation to the structure of spacetime at small scales and its emergence at larger scales \cite{Zeh:1970zz}, and the quantum-to-classical transition \cite{g,x,y,z}, related to the measurement problem, which seeks to explain macroscopic realism \cite{Hornberger}. Although these issues are typically considered separately, Penrose \cite{Schlosshauer:2019ewh} has proposed that gravity might play a role in quantum state reduction, suggesting a possible link between the structure of space-time at small scales and the suppression of quantum effects at larger scales \cite{sb}. 

Various approaches to quantum gravity consistently emphasize the need for a fundamental reevaluation of spacetime, suggesting that it may not be continuous but could exhibit discrete quantum properties at extremely small scales \cite{wn,zz}. This leads to fluctuations commonly referred to as ``quantum spacetime" \cite{zzz}. Achieving absolute precision in localizing events would require probes with such short wavelengths that they demand infinite energy density \cite{doplicher}. However, such extreme conditions could result in gravitational instabilities, such as the formation of black holes and event horizons, which would block communication between observers and the regions of space being examined. Consequently, a perfect localization becomes operationally unattainable, with a Planck length around $10^{-35}$ meters, setting the ultimate limit on precision. This introduces an inherent ``fuzziness" in spacetime, an unavoidable aspect that any theoretical framework for describing phenomena at these minuscule scales must account for \cite{lis1,lis2}.

The concept of space-time ``fuzziness" can be interpreted by modifying the commutator algebra of phase space in quantum mechanics, providing potential insights into the enigmatic nature of quantum gravity phenomenology. However, detecting quantum gravitational effects poses a major challenge, as the energies associated with the Planck scale—roughly 14 orders of magnitude higher than those achievable with current technology, such as the Large Hadron Collider (LHC) at CERN—are far beyond the reach of present-day experiments \cite{zzz}. 

Rather than seeking direct evidence of quantum gravity in the ultraviolet (UV) regime at extremely small scales, recent research has shifted towards exploring its potential effects at larger cosmological scales \cite{lis4}. Particle horizons at these vast distances hint at the possibility of quantum gravitational phenomena manifesting in the infrared (IR) sector, which could become observable at astrophysical or cosmological scales \cite{lis5}. This perspective has gained traction, particularly following the detection of gravitational waves by LIGO \cite{lis6}, which has sparked a new field of multi-messenger astronomy. These advancements have fueled efforts to investigate quantum gravity signatures, including the prospect of probing quantum spacetime through gravitational wave phenomena \cite{s,p,k}.

However, some recent efforts have focused on investigating the observable effects of quantum gravity using experimental data from neutrino sources such as OPERA \cite{op} and MINOS \cite{op1}. Neutrinos, with their unique quantum properties and ability to oscillate between different flavors over long distances, offer a promising pathway for probing otherwise inaccessible aspects of spacetime. Due to their weak interaction with matter, neutrinos serve as ideal probes for exploring the fundamental nature of spacetime across vast distances, making them valuable tools in the search for quantum gravity effects. Notably, neutrino oscillations have already shown that neutrinos possess mass, challenging the Standard Model of particle physics and deepening our understanding of the quantum realm \cite{op2}.

As a physically meaningful application, we focus on \textit{neutrino flavor oscillations}, which are intrinsically quantum mechanical phenomena arising from coherent superpositions of mass eigenstates. The standard treatment of neutrino oscillations assumes that neutrinos remain isolated and maintain quantum coherence during propagation. However, interactions with a stochastic environment---whether conventional or exotic---can disrupt this coherence, leading to \textit{neutrino decoherence}~\cite{lisi,lisi22,25}, which suppresses the characteristic flavor transition probabilities~\cite{25+}.

Of particular interest is the possibility that decoherence in the neutrino sector could reveal imprints of \textit{quantum gravitational effects}~\cite{lisi4, Domi:2024ypm}. Recent experimental studies, particularly those analyzing atmospheric neutrinos at the IceCube Neutrino Observatory, have placed stringent bounds on decoherence parameters that scale \textit{positively with energy} ($\Gamma \propto E^n$, with $n > 0$)~\cite{lisi1,lisi2,lisi3}. These models are motivated by the idea that high-energy neutrinos are more sensitive to Planck-scale fluctuations. In contrast, several other theoretical approaches predict a \textit{negative power-law behavior} ($n < 0$) for the energy dependence of decoherence, where low-energy sector of neutrinos would be more affected~\cite{g1,g2,bd}.

%As a physically meaningful application, we focus on neutrino flavor oscillations, which are intrinsically quantum mechanical phenomena. The study of neutrino oscillations typically assumes that neutrinos remain isolated and preserve quantum coherence during their oscillations. However, interactions with a stochastic environment can disrupt this coherence, leading to neutrino decoherence \cite{lisi,lisi22,25}, which reduces oscillation probabilities \cite{25+}. In particular, recent findings suggest that the detection of decoherence effects within the neutrino sector may reveal a deep connection between neutrinos and quantum gravity \cite{lisi4}. Recent studies have claimed  stringent limits on decoherence parameters that exhibit positive energy dependence ($\Gamma \propto E^n$, where $n > 0$) using data from atmospheric neutrinos detected by the IceCube Neutrino Observatory \cite{lisi1, lisi2, lisi3}. However, there also exist other studies on
%neutrino decoherence, employing different approaches, that indicate a negative
%power law dependence ($n < 0$) \cite{g1,g2,bd}.

In this vein, a key question arises: Could spacetime itself, influenced by quantum-mechanical effects such as non-commutative geometry \cite{td,tod} and Planck-scale fluctuations, contribute to this decoherence? The stochastic fluctuation paradigm of spacetime geometry, aimed at investigating quantum gravity-induced decoherence, has recently been explored \cite{wn,w2}. In fact, the stochastic nature of quantum spacetime introduces an inherent randomness that can be modeled using an open quantum systems approach \cite{fpm}. This framework allows decoherence to be quantified through a Gorini-Kossakowski-Sudarshan-Lindblad (GKSL)-type master equation \cite{w4, wn4}, providing a systematic method for studying the effects of quantum gravity on neutrino oscillations. This formulation provides a robust framework to investigate quantum gravity-induced decoherence, with the non-commutative $\kappa$-Minkowski spacetime offering an ideal setting for a mathematically consistent analysis.

Furthermore, this work investigates neutrino decoherence within the framework of quantum spacetime, specifically utilizing the $\kappa$-Minkowski model \cite{Rev1, Rev, Rev2}. This choice is particularly significant, as the algebra of $\kappa$-Minkowski spacetime emerges as the flat limit of quantum gravity \cite{ori, ori5}. While spacetime remains flat in this model, the induced curvature in momentum space modifies fundamental phase-space relations, potentially leading to deviations from standard quantum field theory. These modifications create a natural setting to explore Planck-scale effects on neutrino decoherence, positioning neutrinos as an ideal probe of quantum gravitational phenomena.

In addition, this study has two primary objectives. First, it employs an open quantum system framework \cite{fpm}, where the neutrino system interacts with an external environment governed by the quantum properties of spacetime. This interaction introduces stochastic fluctuations that affect neutrino coherence, allowing for an investigation into how quantum (non-commutative) spacetime influences neutrino propagation and oscillation probabilities.

Secondly, the study aims to establish a lower bound on the decoherence length, which quantifies the spatial scale at which quantum coherence is effectively lost, marking the transition to classical behavior. Additionally, it examines the energy dependence of the decoherence parameter induced by quantum spacetime fluctuations. A key aspect of this analysis is to distinguish the effects of quantum spacetime from other environmental factors that contribute to neutrino decoherence \cite{ki}. This distinction deepens our understanding of neutrino oscillations in quantum environments and provides insight into the broader interplay between quantum gravity and neutrino physics.

The remainder of this paper is structured as follows. 
Section II derives the Lindblad-type master equation generated by stochastic Planck-scale fluctuations in $\kappa$-Minkowski spacetime. 
In Sec.~III, we apply this framework to three-flavor neutrino oscillations and obtain analytic expressions for the corresponding survival and transition probabilities. 
Section IV examines the characteristic energy dependence of the decoherence parameter and evaluates the minimal coherence length implied by the inverse-power scaling at sub-eV energies, with the cosmic neutrino background serving as a natural infrared benchmark in this context. 
Finally, Sec.~V summarizes our main findings and outlines future directions.

%The remainder of this paper is structured as follows: Section II derives the Lindblad-type master equation within the framework of \(\kappa\)-Minkowski spacetime. In Sec. III, we analyze decoherence in three-flavor neutrino oscillations through the survival and transition probability amplitudes. Section IV investigates the energy dependence of the decoherence parameter due to the stochastic behavior associated with quantum spacetime, specifically \(\kappa\)-Minkowski. We also estimate the minimal coherence length of the oscillation using the cosmic neutrino background ($C\nu B$) within the framework of stochastic fluctuations in quantum spacetime. Lastly, Sec. V summarizes our conclusions.

\section{Quantum space time and Lindblad
master equation }

\subsection{Basic Overview of \texorpdfstring{\(\kappa\)}{kappa} Minkowski-Type Quantum Spacetime}

Here we provide a brief overview of the key features of the quantum nature of spacetime in the context of the flat limit of quantum gravity \cite{ori}. In a (2+1) dimensional model with a nonzero cosmological constant \(\Lambda\), the classical symmetry group becomes the de Sitter or Anti-de Sitter group, \(SO(3,1;\Lambda)\) or \(SO(2,2;\Lambda)\), depending on the sign of \(\Lambda\), and reduces to the Poincaré group as \(\Lambda \rightarrow 0\). At the quantum level, this symmetry is described by the quantum-deformed groups \(SO_q(3,1;\Lambda)\) or \(SO_q(2,2;\Lambda)\), with \(q = \exp(-l_P \Lambda)\), where \(l_P\) is the Planck length \cite{ori1}. For small \(\Lambda\), \(q\) approaches 1, recovering the classical group. In the flat limit \(\Lambda l_P \rightarrow 0\), the symmetry contracts to the \(\kappa\)-deformed Poincaré group, suggesting it as the symmetry of weak quantum gravity in flat space-time \cite{Rev1}.

In (3+1) dimensions, it is conjectured that the vacuum symmetry at \(\Lambda \neq 0\) is described by a \(q\) deformed de Sitter or Anti-de Sitter group, which also contracts to the \(\kappa\)-Poincaré algebra in the flat limit \cite{luk}. This indicates that \(\kappa\)-Poincaré symmetry governs perturbations around the classical Minkowski spacetime in quantum gravity \cite{ori2, ori3}. Born’s reciprocity principle \cite{o} further implies that a small cosmological constant (that is, a flat limit of the curved spacetime) and maintaining the curvature of the momentum space lead to the flat limit of quantum gravity and, particularly, to the \(\kappa\)-deformed Poincaré flat quantum spacetime, which has recently been explored in \cite{ori5,ori7}.

We explore the \(\kappa\)-Minkowski quantum spacetime, characterized by the commutation relation
\begin{equation} 
[\hat{x}_{\mu}, \hat{x}_{\nu}] = i(a_{\mu}\hat{x}_{\nu} - a_{\nu}\hat{x}_{\mu}),
\label{kappa}
\end{equation}
where \(\hat{x}_{\mu}\) are non-commutative space-time coordinates and ($a_{\mu}$) is a set of four constants, which are real scalars and can be identified with the set of four deformation parameters. In the limit \(a_{\mu} \rightarrow 0\), the spacetime becomes commutative, recovering the standard flat Minkowski space. This framework provides a natural description of quantum spacetime in the language of non-commutative geometry \cite{ori8}. 

To express \(\hat{x}_{\mu}\) in terms of commutative auxiliary coordinates \(q_{\mu}\) and their canonical momenta \(p_{\mu}\), which satisfy the relations  
\begin{equation}
[q_{\mu}, q_{\nu}] = 0, \quad [q_{\mu}, p_{\nu}] = -i\hbar \eta_{\mu\nu}, \quad [p_{\mu}, p_{\nu}] = 0,  
\end{equation}  
where \(\eta_{\mu\nu} = \text{diag}(+1, -1, -1, -1)\), we adopt a perturbative approach. This involves expanding \(\hat{x}_{\mu}\) to a Hermitian realization up to first order in the deformation parameter \(a_{\mu}\)~\cite{thermo,hri}:  
\begin{equation}\label{xhat}
q_{\mu}\rightarrow \hat{x}_{\mu} = q_{\mu} + \delta q_{\mu}(a),
\end{equation}  
where the correction term \(\delta q_{\mu}(a)\) is expressed as  
\begin{equation}\label{varx}
\delta q_{\mu}(a) = \frac{1}{2}[\alpha \frac{(a \cdot p)}{\hbar} q_{\mu} +  \beta\frac{(a \cdot q)}{\hbar}p_{\mu} + \text{h.c.}],\quad \alpha, \beta \in \mathbb{R},
\end{equation}  
and h.c.\ denotes the Hermitian conjugate.  

To maintain the commutation relation~\eqref{kappa} up to first order in \(a_{\mu}\), the parameter \(\alpha\) must satisfy the condition \(\alpha = -1\), while \(\beta\) remains arbitrary, and will be fixed from a phenomenological perspective. It is important to note that the coordinate transformations in~\eqref{xhat} are defined using symmetric ordering. However, by adopting alternative operator orderings, these transformations can be expressed in a manifestly Hermitian form, leading to equivalent theoretical descriptions~\cite{th, thk}.

Next, we turn to the task of determining the deformed momentum operator \(\hat{p}_{\mu}\). Following \textit{Feynman’s prescription} \cite{hri1,hri2}, we begin with the canonical momentum operator in the commutative case, which is expressed as \(p_{\mu} = m\frac{d x_{\mu}}{d \tau}\), with \(\frac{dp_{\mu}}{d\tau}=0\). Utilizing Eq. (\ref{xhat}), we can then write
\begin{equation}
\hat{p}_{\mu} = p_{\alpha}E^{\alpha} \! _{\mu} \rightarrow  p_{\mu} + \delta\hat{p}_{\mu}(a),
\end{equation}
and require that the commutation relations and Jacobi identities between \(\hat{p}_{\mu}\), \(\hat{x}_{\nu}\), and \(\hat{x}_{\rho}\) are satisfied up to the first order in \(a\). Substituting \(x\) with \(p\) (as \(\dot{p}_{\mu} = 0\)), one finds
\begin{equation}\label{p-nonc}
\delta\hat{p}_{\mu}(a) = \frac{(\alpha+\beta)}{\hbar}(a \cdot p)p_{\mu} .
\end{equation}
Thus, the commutation relation is given by
\begin{equation}\label{comm-NC}
[\hat{p}_{\mu}, \hat{x}_{\nu}] = i\eta_{\mu\nu}(\hbar + s(a \cdot p)) + i(s+2)a_{\mu}p_{\nu} + i(s+1)a_{\nu}p_{\mu},
\end{equation}
where \(s = 2\alpha+\beta\).

Considering a relativistic particle in \(\kappa\)-deformed space, the Hamiltonian is given by
\begin{equation}
\hat{H} =\hat{p}_{0}c= \sqrt{\hat{p}^2c^{2} + m^2c^{4}}.
\label{s}
\end{equation}
Here,
\begin{equation}
\hat{p} = \tilde{p}\left(1 + (\alpha + \beta)\frac{(a \cdot p)}{\hbar}\right)
\label{4}
\end{equation}
is the deformed spatial momentum represented in terms of the usual canonical momentum operators, and $\tilde{p}=p_i^2$. In the ultra-relativistic limit, (\ref{s}) may be approximated as
\begin{equation}
\hat{H} \approx \hat{p}c + \frac{m^2c^{3}}{2\hat{p}}.
\label{a}
\end{equation}
The spatial part of the deformation parameter may be eliminated assuming a time-like deformation \(a_\mu = (a_{0}, 0, 0, 0)\), which in turn gives  
\begin{equation}
    \hat{p}_{0}=p_{0}+\frac{(\alpha+\beta)}{\hbar} a_{0}p^{2}_{0}.
    \label{x}
\end{equation}
By now substituting Eqs. (\ref{4}), and (\ref{x}) into (\ref{a}) and self-consistently solving for the physical root of \( p_0 \), which corresponds to the magnitude of the standard commutative momentum, we obtain the commutative equivalent of the Hamiltonian from (\ref{x}), leading to the result
\begin{equation}
\hat{H}_{\text{eff}} = \left[ \tilde{p} c + \frac{m^2 c^3}{2 \tilde{p}} \right] + (\alpha + \beta) \frac{a_0 c}{\hbar} \tilde{p}^2 + \mathcal{O}(a_0^2, m^4),
\end{equation}
with $\alpha+\beta\neq 0$.

Using this Hamiltonian in the Schrödinger equation, and omitting terms of order $a^2_0$ and $m^4$ and higher, one gets
\begin{equation}
i\hbar\frac{d|\psi\rangle}{dt} = \hat{H}_{\text{eff}}|\psi\rangle = \left(\hat{H}_{0} + \hat{H}_{\text{int}}\right)|\psi\rangle,
\end{equation}
where
\begin{equation}
\hat{H}_{0} = \left[\tilde{p}c + \frac{m^2c^{3}}{2\tilde{p}}\right], \quad \hat{H}_{\text{int}} = (s+1)\frac{a_{0}c}{\hbar} \tilde{p}^2.
\end{equation}
Here, we used the fact that \( \alpha = -1 \), as previously noted from the consistency of the spacetime coordinate algebra (\ref{kappa}), and identified \( \alpha + \beta = s + 1 \).

In the interaction picture, the state $|\psi\rangle$ is expressed as
\begin{equation}
|\psi(t)\rangle_{I} = \exp{\left(\frac{i}{\hbar}H_{0}t\right)}|\psi(t)\rangle_{S},
\end{equation}
such that the Schrödinger equation in this representation is
\begin{equation}
i\hbar\frac{\partial |\psi(t)\rangle_{I}}{\partial t} = \hat{H}^{I}_{\text{int}}(t)|\psi(t)\rangle_{I},
\end{equation}
where
\[
\hat{H}^{I}_{\text{int}}(t) = \exp{\left(\frac{iH_{0}t}{\hbar}\right)}\hat{H}_{\text{int}}\exp{\left(\frac{-iH_{0}t}{\hbar}\right)}.
\]
For convenience let us rewrite the interaction Hamiltonian $\hat{H}^I_{\rm int}(t)$ in the form 
\begin{equation}
	\hat{H}^I_{\rm int}(t) = \lambda a_0\hat{V}(t),
\end{equation}
where $\lambda$ is a dimensionless bookkeeping parameter introduced to
systematically organize the perturbative expansion of the dynamical map.
At the end of the calculation, $\lambda$ will be absorbed into the
coupling constant $a_0$. The operator $\hat{V}(t)$ is defined as
\begin{equation}
\hat{V}(t) = \frac{(s+1)c}{\hbar}\,\tilde{p}^{\,2}_{I}(t).
\end{equation}
Therefore, the Liouville-von Neumann equation becomes
\begin{equation}
\frac{\partial\hat{\varrho}_{I}}{\partial t} = -\frac{i\lambda a_0}{\hbar}[\hat{V}(t), \hat{\varrho}_{I}] := \mathcal{L}_I(t)\hat{\varrho}_I.
\end{equation}
The formal solution to this equation can be constructed by integrating both sides and iteratively inserting the result back into the righthand side: 
\begin{equation}\label{myeq4}
	\hat{\varrho}_I(t) = \Lambda(t)\hat{\varrho}_I(0).
\end{equation}
Here,
\begin{align}
	 \Lambda(t) = \mathcal{I} + &\sum^{\infty}_{n=1}(-i\lambda)^n\int^{t}_0dt_n\int^{t_n}_0dt_{n-1}... \nonumber\\
    &\times\int^{t_2}_0dt_1\mathcal{L}_I(t_n)...\mathcal{L}_I(t_1),
\end{align}
and $\mathcal{I}$ is the identity superoperator.

\subsection{Quantum spacetime with stochastic fluctuations}

In the present framework, the quantum concept of spacetime can undergo random fluctuations, leading to a probabilistic description of its geometry. This can be incorporated by treating the deformation parameter \(a_{0}\) as a stochastic parameter (c-number) that represents fluctuations in spacetime at the Planck scale, as suggested by various models of quantum gravity \cite{dis, to, disquant}. To capture the inherent randomness of these fluctuations, we model this parameter as a stochastic process with a short correlation time 
$\tau\sim t_p$. In the limit where the system dynamics cannot resolve this scale 
($t_p\ll t_{\rm sys}$), the noise may be treated as Gaussian white noise, 
characterized by a zero mean and an effective autocorrelation function:

 %we model this parameter as Gaussian white noise, characterized by a zero mean and a well-defined autocorrelation function:
\begin{equation}
    a_{0} \rightarrow a_{0}(t) = a_0 \sqrt{t_p} \, \chi_{0}(t),
\end{equation}
with
\begin{equation}
 \langle \chi_{0}(t) \rangle = 0, \quad  \langle \chi_{0}(t) \chi_{0}(t') \rangle = \delta(t - t').
 \label{chi_stoch}
\end{equation}
Here, the parameter \( a_0 \) is a dimensionful quantity, typically associated with the Planck length scale (\( a_0 \sim \ell_p \)), where \( t_p \) denotes the Planck time. The notation \( \langle \cdot \rangle \) represents an average over fluctuations.

The assumption of a zero mean for the noise term \(\chi_{0}(t)\) reflects the absence of inherent bias in spacetime deformations, aligning with the premise that classical spacetime is commutative on average. The instantaneous amplitude \(\chi_{0}(t)\) is specified to be of the order \(\frac{1}{\sqrt{t_p}}\), ensuring that the deformation parameter \(a_0(t)\) fluctuates at the Planck scale:
\begin{equation}
a_0(t) \sim a_0 \sqrt{t_p} \cdot \frac{1}{\sqrt{t_p}} = a_0.
\end{equation}
To proceed, we assume that the noise term \( \chi_0(t) \) undergoes fluctuations on a characteristic timescale \( \tau \sim t_p \), intrinsic to spacetime's quantum structure. These rapid fluctuations are averaged over a much longer system timescale \( t_{\text{sys}} = \hbar / E_0 \), associated with the energy scale of the neutrino. In particular, with $t_p\simeq 5.39\times 10^{-44}{\rm s}$, the ratio of the two timescales is 
 \begin{equation}
 	\frac{t_p}{t_{\rm sys}} \sim \frac{t_pE}{\hbar} \simeq 8.2\times 10^{-29}E[{\rm eV}].
    \label{x}
 \end{equation}
For C$\nu$B neutrinos, $E\sim 10^{-6}{\rm eV}$, so that $t_p/t_{\rm sys}\sim10^{-35}$, whereas for TeV neutrinos, $t_p/t_{\rm sys}\sim10^{-17}$. Since $t_p\ll t_{\rm sys}$, the time derivative $\dot{a}_0(t)$ is thus suppressed by a factor $\tau/t_{\rm sys}\ll1$ and may be neglected in the dynamical equations.

Accordingly, while the deformation parameter \( a_0(t) \) is modeled as a stochastic process, it may be treated as effectively constant on system timescales. This approximation allows us to retain the consistency of the noncommutative algebra of spacetime coordinates 
in its standard form ~(\ref{kappa}) even with a time-dependent stochastic \( a_0(t) \).

While this framework captures the essential features of decoherence induced by quantum spacetime fluctuations, further generalizations—such as treating time-dependent deformation parameters dynamically—may provide complementary perspectives. In this context, recent works have explored the emergence of noncommutativity from stochastic processes~\cite{w2}, the role of fluctuating minimal length in quantum gravitational decoherence~\cite{gwt}.

Averaging over the noise in the evolution equation for the statistical operator $\hat{\varrho}_I(t)$ in Eq. \eqref{myeq4} now yields 
\begin{equation}
	\langle\hat{\varrho}_I(t)\rangle = \langle\Lambda(t)\rangle\hat{\varrho}_I(0). 
\end{equation}
If the map $\Lambda(t)$ is assumed to be invertible, then the time derivative of the above can be expressed as 
\begin{equation}
	\left\langle\frac{\partial\hat{\varrho}_I}{\partial t}\right\rangle = \left\langle\frac{\partial\Lambda(t)}{\partial t}\right\rangle\langle\Lambda^{-1}(t)\rangle\langle\hat{\varrho}_I(t)\rangle := \mathcal{K}(t)\langle\hat{\varrho}_I(t)\rangle,
\end{equation}
where $\mathcal{K}(t)$ defines the time-local generator of the resulting master equation. To solve this equation, it is necessary to follow a perturbative approach by expanding the generator $\mathcal{K}(t)$ in powers of the coupling parameter $\lambda$. As shown rigorously in \cite{d2,fpm}, such an expansion can be written as
\begin{align}\label{myeq5}
	\mathcal{K}(t) = &\sum^{\infty}_{n=1}\lambda^n\int^t_0dt_n\int^{t_n}_0dt_{n-1}... \nonumber\\
    &\times\int^{t_2}_0dt_1\langle\langle\mathcal{L}_I(t)\mathcal{L}_I(t_n)...\mathcal{L}_I(t_1)\mathcal \rangle\rangle,
\end{align}
where $\langle\langle\cdot\rangle\rangle$ represent time-ordered cumulants. Under the Born approximation we assume the series in Eq. \eqref{myeq5} can be truncated to second-order in $\lambda$, such that
\begin{equation}
	\mathcal{K}(t) \approx \lambda\langle\mathcal{L}_I(t)\rangle + \lambda^2\int^t_0dt'\,\langle\langle\mathcal{L}_I(t)\mathcal{L}_I(t-t')\rangle\rangle,
\end{equation}
with $\langle\langle\mathcal{L}_I(t)\mathcal{L}_I(t-t')\rangle\rangle = \langle\mathcal{L}_I(t)\mathcal{L}_I(t-t')\rangle - \langle\mathcal{L}_I(t)\rangle\langle\mathcal{L}_I(t-t')\rangle$. The terms proportional to $\langle\mathcal{L}_I(t)\rangle$ will vanish since the stochastic deformation parameter has zero mean. Hence, for sufficiently weak noise $\chi_0(t)$, the above master equation simplifies to 
\begin{align}
	\left\langle\frac{\partial\hat{\varrho}_I}{\partial t}\right\rangle &\approx -\lambda^2\frac{a^2_0t_p}{\hbar^2}\int^t_0ds\langle \chi_0(t)\chi_0(t')\rangle[\hat{V}(t),[\hat{V}(t'),\langle\hat{\varrho}_I(t)\rangle]] \nonumber\\
						       &= -\lambda^2\frac{a^2_0t_p}{\hbar^2}[\hat{V}(t),[\hat{V}(t),\langle\hat{\varrho}_I(t)\rangle]].
\end{align}

Here we have used Eq.~\eqref{chi_stoch}, where the $\delta$ correlation is to be interpreted as an effective, coarse-grained description of Planck-scale fluctuations. At the microscopic level the noise correlator has a finite correlation time of order $t_p$, which reduces to a Dirac delta in the Markovian limit $t_p \ll t_{\rm sys}$ (\ref{x}), as is standard in open quantum systems theory~\cite{fpm}.

By now introducing the simplified notation $\hat{\rho}_I(t):=\langle\hat{\varrho}_I(t)\rangle$ and absorbing $\lambda$ into the coupling parameter $a_0$, we obtain 
\begin{equation}
\frac{d\hat{\rho}_{I}}{dt} = -\sigma_{0}
\left[ \tilde{p}^2(t),[ \tilde{p}^2(t), \hat{\rho}_{I}(t)]\right],
\end{equation}  
where \(\sigma_0 = \frac{c^2 \chi t_p a_0^2}{\hbar^4}\) and \(\chi = (s+1)^2\) encapsulate the stochastic corrections originating from Planck-scale fluctuations.

Finally, the master equation in the Schr\"{o}dinger picture reads
\begin{equation}
\frac{d\hat{\rho}_{s}}{dt} = \, \frac{-i}{\hbar} \left[
\left(
\tilde{p} c + \frac{m c^{3}}{2 \tilde{p}}
\right), \hat{\rho}_{s}(t)
\right] - \sigma_{0} \left[
\tilde{p}^2 , 
[
\tilde{p}^2, \hat{\rho}_{s}(t)
]
\right].
\label{myeq6}
\end{equation}
Expressing this in terms of the unperturbed Hamiltonian \(H_0\), up to the leading-order mass correction, we have
\begin{align}
\frac{d\rho_s}{dt} = & -\frac{i}{\hbar}[H_{0}, \rho_s(t)] \nonumber \\
& - \sigma \left[\left(H^{2}_{0} -m^{2}c^{4}\right), \left[\left(H^{2}_{0} - m^{2}c^{4}\right), \rho_s(t)\right]\right],
\end{align}
with $\sigma=\frac{\sigma_{0}}{c^{4}}$. This equation can be identified as a Gorini-Kossakowski-Sudarshan-Lindblad (GKSL) equation \cite{w4, wn4, wn5, g5} and expressed as follows:
\begin{equation}
\frac{d\rho}{dt} = -i[H_{0}, \rho(t)] - [D, [D, \rho(t)]],
\label{myeq}
\end{equation}
where \( D = \sqrt{\sigma} \left(H^{2}_{0} - m^{2}c^{4} \right) \) defines the corresponding Lindblad operator, considering only the mass-dependent term of the leading order. Naturally, the first term in Eq. (\ref{myeq}) governs the unitary evolution of the system, while the second term introduces a non-unitary contribution that drives the decoherence dynamics.

\section{Decoherence From Quantum Space time}

\subsection{Quantum Mechanics of Neutrino Oscillations in Three Flavors}

We first provide a brief discussion of the framework for neutrino oscillations through standard quantum mechanics principles without
having to take into account considerations of the fermionic nature of neutrinos \cite{fermion,pbp}. 
Let us represent the neutrino states with masses \( m_i \) (where \( i = 1, 2, 3 \)) and momentum \( p_{0} \) by \( \ket{\nu_i} \). These states are treated as eigenstates of the free Hamiltonian, \( \hat{H} \), such that they satisfy
\begin{equation}
    \hat{H}\ket{\nu_i} = E_i(\tilde{p}_{0})\ket{\nu_i},
\end{equation}
where $E_i(\tilde{p}_{0}) \approx E + \frac{m_i^2c^{4}}{2E}$ and $\tilde{p}_{0}c=E$.

Next, we define the corresponding flavor states \( \ket{\nu_A} \) (for \( A = e, \mu, \tau \)), corresponding to the electron, muon, and tau neutrinos, in terms of the mass states through the Pontecorvo–Maki–Nakagawa–Sakata (PMNS) matrix \( U_{A j} \) as
\begin{equation}
    \ket{\nu_A} = \sum_j U_{A j} \ket{\nu_j},
    \label{hj}
\end{equation}
where the PMNS matrix satisfies the following conditions:
\begin{equation}
    \sum_j U_{A j} U_{B j}^* = \delta_{AB}, \quad
    \sum_A U_{A j} U_{A k}^* = \delta_{jk}.
\end{equation}
 The PMNS matrix \cite{pbp1} can be written as
\footnotesize
\begin{equation}
    U = \begin{pmatrix}
         c_{12}c_{13} & s_{12}c_{13} & s_{13}e^{i\delta} \\
         -s_{12}c_{23} - c_{12}s_{23}s_{13}e^{i\delta} & c_{12}c_{23} - s_{12}s_{23}s_{13}e^{i\delta} & s_{23}c_{13} \\
         s_{12}s_{23} - c_{12}c_{23}s_{13}e^{i\delta} & -c_{12}s_{23} - s_{12}c_{23}s_{13}e^{i\delta} & c_{23}c_{13}
    \end{pmatrix},
    \label{pmns}
\end{equation}
\normalsize
where \( c_{ij} = \cos\theta_{ij} \) and \( s_{ij} = \sin\theta_{ij} \). Here, \( \theta_{ij} \) are the mixing angles, \( \delta \) is the CP-violating phase, and Majorana phases are ignored as we are working with Dirac neutrinos. The mixing angles \( \theta_{ij} \) are taken in the first quadrant, and \( \delta \) range between 0 and \( 2\pi \).
The matrix \( U \) can be factored into a product of rotation matrices \( \mathcal{O}_{ij} \), each representing a rotation in the \( ij \)-plane \cite{akhmedov}
\begin{equation}
    U = \mathcal{O}_{23} \mathcal{U}_{\delta} \mathcal{O}_{13} \mathcal{U}_{\delta}^\dagger \mathcal{O}_{12},
\end{equation}
where \( \mathcal{U}_{\delta} = \text{diag}(1, 1, e^{i\delta}) \).

The Schrödinger equation for neutrino states in the mass basis, considering the three-flavor scenario and disregarding an additional constant shift proportional to the identity matrix \cite{akh, kh}, can be expressed as
\begin{equation}
i\hbar\frac{d}{dt} \begin{pmatrix} \nu_1 \\ \nu_2 \\ \nu_3 \end{pmatrix} = \left[ \frac{c^{4}}{2E} \begin{pmatrix} m_1^2 & 0 & 0 \\ 0 & m_2^2 & 0 \\ 0 & 0 & m_3^2 \end{pmatrix} \right] \begin{pmatrix} \nu_1 \\ \nu_2 \\ \nu_3 \end{pmatrix}.
\end{equation}
In the three-flavor case, there are two possible mass hierarchies: the normal hierarchy and the inverted hierarchy. For the normal mass hierarchy, we assume
\begin{equation}
m_3^2 \gg m_2^2 > m_1^2.
\end{equation}
Thus, in the flavor basis, the Schrödinger equation becomes
\begin{equation}
i\hbar\begin{pmatrix} \dot{\nu_e} \\ \dot{\nu_\mu} \\ \dot{\nu_\tau} \end{pmatrix} = U\left[\frac{1}{2E} \begin{pmatrix} 0 & 0 & 0 \\ 0 & \Delta m_{21}^2 c^{4} & 0 \\ 0 & 0 & \Delta m_{31}^2 c^{4} \end{pmatrix}  \right] U^{\dagger} \begin{pmatrix} \nu_e \\ \nu_\mu \\ \nu_\tau \end{pmatrix},
\label{g}
\end{equation}
where \( \Delta m_{ij}^2 = m_i^2 - m_j^2 \). The terms proportional to the identity matrix contribute only to a global phase, which does not impact the oscillation probabilities.

\subsection{Decoherence of neutrino oscillations in \texorpdfstring{\(\kappa\)}{kappa} Minkowski spacetime}

The GKSL equation (\ref{myeq}) provides a useful framework for analyzing decoherence patterns in three-flavor neutrino oscillations. The GKSL equation (\ref{myeq}) provides a useful framework for analyzing
decoherence patterns in three-flavor neutrino oscillations.
We note that the emergence of the GKLS structure does not rely on the
environment being quantum: in the present case it arises from averaging over
classical stochastic spacetime fluctuations in the Markovian limit, which
guarantees complete positivity and trace preservation of the reduced dynamics,
as is well known for systems driven by classical stochastic Hamiltonians
\cite{gwt,Rivas2012}.

The effective unperturbed Hamiltonian in the mass eigenbasis can be read off from (\ref{g}) and is given as follows:
\begin{equation}
    H_0 = \frac{1}{2E} \begin{pmatrix} 0 & 0 & 0 \\ 0 & \Delta m_{21}^2c^{4} & 0 \\ 0 & 0 & \Delta m_{31}^2c^{4} \end{pmatrix}.
\label{myeq10}\end{equation}
As a result, the Lindblad operators in the mass eigenbasis for three flavors can be expressed as follows:
\begin{align}
    D_{m} = \sqrt{\sigma}\,&\textrm{diag}(-m_{1}^2 c^{4}, \, \omega^2 (\Delta m_{21}^2c^{4})^{2} - m_2^2 c^{4}, \nonumber \\
    &\omega^2 (\Delta m_{31}^2c^{4})^{2} - m_3^2 c^{4})
\label{myeq11}
\end{align}
where \( \omega = \frac{1}{2E} \). The general Hermitian form of the density matrix for the three-flavor case in the mass basis representation is given by
\begin{equation}
    \rho_{m}(t) = \begin{pmatrix} 
    a & p + iq & f + ig \\ 
    p - iq & b & x + iy \\ 
    f - ig & x - iy & c 
    \end{pmatrix},
\label{myeq12}
\end{equation}
where the parameters \( a, b, c, p, q, f, g, x, y \) are real functions of time, and it is understood that the trace of the matrix satisfies \( \text{Tr}(\rho) = a + b + c = 1 \). Using Eqs. (\ref{myeq10}), (\ref{myeq11}), and (\ref{myeq12}) in the Lindblad equation, one obtains
\begin{equation}
    \begin{pmatrix}
        \dot{a} & \dot{p} + i \dot{q} & \dot{f} + i \dot{g} \\
        \dot{p} - i \dot{q} & \dot{b} & \dot{x} + i \dot{y} \\
        \dot{f} - i \dot{g} & \dot{x} - i \dot{y} & \dot{c}
    \end{pmatrix} = \begin{pmatrix}
        A_{11} & A_{12} & A_{13} \\
        A_{12}^\star & A_{22} & A_{23} \\
        A_{13}^\star & A_{23}^\star & A_{33}
    \end{pmatrix}.
\end{equation}
It is worthwhile noting that, if we consider the inverted hierarchy, i.e., $m^2_2 > m^2_1 \gg m^2_3$, the Lindblad equation remains unchanged. Given also that $D_m$ is diagonal in the mass eigenbasis, only the off-diagonal
elements of $\rho_m(t)$ will evolve in time, i.e. the Lindblad equation in this
basis generates a pure dephasing dynamics. In this mass (energy) eigenbasis,
relaxation—understood as population transfer between eigenstates—is absent,
while decoherence appears solely as dephasing of the off-diagonal density-matrix
elements.
  Hence,
\begin{equation} 
A_{11} = A_{22} = A_{33} = 0,
\end{equation}
while the off-diagonal elements take the following forms (in natural units $\hbar=c=1$):
\begin{equation}
    \begin{aligned}
    A_{12} &= \left[ -\omega \Delta m_{21}^2 q - \sigma \omega^4 (\Delta m_{21}^2)^4 p \right] \\
    &\quad + i \left[ \omega \Delta m_{21}^2 p - \sigma \omega^4 (\Delta m_{21}^2)^4 q \right],
    \end{aligned}
\end{equation}
\begin{equation}
    \begin{aligned}
    A_{13} &= \left[ -\omega \Delta m_{31}^2 g - \sigma \omega^4 (\Delta m_{31}^2)^4 f \right] \\
    &\quad + i \left[ \omega \Delta m_{31}^2 f - \sigma \omega^4 (\Delta m_{31}^2)^4 g \right],
    \end{aligned}
\end{equation}
\begin{align*}
A_{23} &= \left[ -\omega \Delta m_{32}^2 y - \sigma \omega^4 (\Delta m_{32}^2)^2 (\Delta m_{31}^2 + \Delta m_{21}^2)^2 x \right] \\
&\quad + i \left[ \omega \Delta m_{32}^2 x - \sigma \omega^4 (\Delta m_{32}^2)^2 (\Delta m_{31}^2 + \Delta m_{21}^2)^2 y \right].
\end{align*}
Since the diagonal elements of the density matrix \(\rho\) are time independent, one has
\begin{equation}
    a(t) = a(0), \quad b(t) = b(0), \quad c(t) = c(0).
\end{equation}

Using the Hamiltonian, Lindblad operators, and the density matrix to express the Lindblad equation, the time evolution of the off-diagonal elements are
given by
\begin{equation}
    \begin{aligned}
    p(t) &= e^{-\sigma \omega^4(\Delta m^2_{21})^4 t} \times \left[ p(0) \cos{\left(\omega \Delta m^2_{21} t\right)} \right. \\
    &\quad \left. - q(0) \sin{\left(\omega \Delta m^2_{21} t\right)} \right],
    \end{aligned}
\end{equation}
\begin{equation}
    \begin{aligned}
    q(t) &= e^{-\sigma \omega^4(\Delta m^2_{21})^4 t} \times \left[ q(0) \cos{\left(\omega \Delta m^2_{21} t\right)} \right. \\
    &\quad \left. + p(0) \sin{\left(\omega \Delta m^2_{21} t\right)} \right],
    \end{aligned}
\end{equation}
\begin{equation}
    \begin{aligned}
     f(t) &= e^{-\sigma \omega^4(\Delta m^2_{31})^4 t} \times \left[ f(0) \cos{\left(\omega \Delta m^2_{31} t\right)} \right. \\
     &\quad \left. - g(0) \sin{\left(\omega \Delta m^2_{31} t\right)} \right],
    \end{aligned}
\end{equation}
\begin{equation}
    \begin{aligned}
     g(t) &= e^{-\sigma \omega^4(\Delta m^2_{31})^4 t} \times \left[ g(0) \cos{\left(\omega \Delta m^2_{31} t\right)} \right. \\
     &\quad \left. + f(0) \sin{\left(\omega \Delta m^2_{31} t\right)} \right],
    \end{aligned}
\end{equation}
\begin{equation}
    \begin{aligned}
    x(t) &= e^{-\sigma \omega^4(\Delta m^2_{32})^2 (\Delta m^2_{31} + \Delta m^2_{21})^2 t} \times \left[ x(0) \cos{\left(\omega \Delta m^2_{32} t\right)} \right. \\
    &\quad \left. - y(0) \sin{\left(\omega \Delta m^2_{32} t\right)} \right],
    \end{aligned}
\end{equation}
\begin{equation}
\begin{aligned}
    y(t) &=e^{-\sigma \omega^4(\Delta m^2_{32})^2 (\Delta m^2_{31}+\Delta m^2_{21})^2 t} \times \left[y(0) \cos{\left(\omega \Delta m^2_{32}t\right)} \right. \\ &\quad \left.+x(0)\sin{\left(\omega \Delta m^2_{32}t\right)} \right].
    \end{aligned}
\end{equation}
To determine the parameters of the initial density matrix $\rho_m(0)$ in terms of those from Eq. (\ref{pmns}), the following transformation is carried out from the mass basis to the flavor basis
\begin{equation}  
    \rho_{A}(t) = U \rho_{m}(t) U^{\dagger},  
\end{equation}  
which is evaluated at \( t = 0 \). Comparing this with the initial flavor state  
\begin{equation}  
    \rho_A(0) = \sum_{i,j} U_{Ai} U_{Aj}^* \left| \nu_i \right\rangle \left\langle \nu_j \right|  
\end{equation}  
yields the desired parameters. It should be noted that in this basis the same dephasing dynamics described by Eq. \eqref{myeq} now manifests as damping of oscillation amplitudes. The oscillation probabilities for each channel can be calculated from
\begin{equation}
P(\nu_A\rightarrow \nu_B;t) = \text{Tr} \left[ \rho_A(t) \rho_B(0) \right].
\end{equation}
Thus, the survival and transition probabilities for electron and muon neutrinos, incorporating decoherence effects, can be derived using this formalism. Such probabilities read
\begin{equation}
\begin{aligned}
 P(\nu_e \to \nu_e&; t) = \text{Tr} \left( \rho_{e}(t) \, |\nu_e \rangle \langle \nu_e| \right) = \\ 
 &\left( 1 - 2c^2_{12} s^2_{12} c^4_{13} - 2c^2_{12} c^2_{13} s^2_{13} - 2s^2_{12} c^2_{13} s^2_{13} \right) \\ 
 &+ 2 \cos\left(\frac{\Delta m^2_{21} t}{2E} \right) e^{-\sigma \omega^4 (\Delta m^2_{21})^4 t} c^2_{12} s^2_{12} c^4_{13} \\ 
 & + 2 \cos\left(\frac{\Delta m^2_{31} t}{2E} \right) e^{-\sigma \omega^4 (\Delta m^2_{31})^4 t} c^2_{12} c^2_{13} s^2_{13} \\ 
 & + 2 \cos\left(\frac{\Delta m^2_{32} t}{2E} \right) \times\\ 
 & e^{-\sigma \omega^4 (\Delta m^2_{32})^2 (\Delta m^2_{31} + \Delta m^2_{21})^2 t}  s^2_{12} c^2_{13} s^2_{13},
\end{aligned}
\label{l}
\end{equation}
\begin{equation}
    \begin{aligned}
      P(e \to \mu&; t) = \text{Tr} \left( \rho_{e}(t) \, |\nu_\mu \rangle \langle \nu_\mu| \right) = \\ 
 & 2c^2_{12} s^2_{12} c^2_{13}(c^2_{23} - s^2_{13} s^2_{23}) - 2s^2_{23} c^2_{13} s^2_{13} \\ 
        & - 2\cos{\delta}(c_{12} s^3_{12} c_{23} s_{23} c^2_{13} s_{13} - c^3_{12} s_{12} c_{23} c_{13} s_{13}) \\ 
        & - 4 \{ c^2_{12} s^2_{12} c^2_{13}(c^2_{23} - s^2_{23} s^2_{13}) \\ 
        & - s_{13} c_{23} c^2_{13} \cos\delta (c_{12} s^3_{12} s_{23} s_{23} - c^3_{12} s_{12}) \} \\ 
        & \times \cos\left(\frac{\Delta m^2_{21} t}{2E}\right) e^{-\sigma \omega^4 (\Delta m^2_{21})^4 t} \\ 
        & + 2(c_{12} s_{12} c_{23} s_{23} c^2_{13} s_{13} \cos\delta + c^2_{12} s^2_{23} c^2_{13} s^2_{13}) \\ 
        & \times \cos\left(\frac{\Delta m^2_{31} t}{2E}\right) e^{-\sigma \omega^4 (\Delta m^2_{31})^4 t} \\ 
        & + 4c^2_{13}(s^2_{12} s^2_{23} s^2_{13} - c_{12} s_{12} c_{23} s_{23} s_{13} \cos\delta) \\ 
        & \times \cos\left(\frac{\Delta m^2_{32} t}{2E}\right) e^{-\sigma \omega^4 (\Delta m^2_{32})^2 (\Delta m^2_{31}+\Delta m^2_{21})^2 t} \\ 
        & + 2c^2_{13}(c_{12} s^3_{12} c_{23} s_{23} s_{13} \sin\delta + c^3_{12} s_{12} c_{23} s_{23} s_{13} \\ 
        & \sin\delta) \times \sin\left(\frac{\Delta m^2_{21} t}{2E}\right) e^{-\sigma \omega^4 (\Delta m^2_{21})^4 t} \\ 
        & + 2c_{12} s_{12} c_{23} s_{23} c^2_{13} s_{13} \sin\delta \sin\left(\frac{\Delta m^2_{31} t}{2E}\right) \\ 
        & \times e^{-\sigma \omega^4 (\Delta m^2_{31})^4 t} \\ 
        & + 2c_{12} s_{12} c_{23} s_{23} c^2_{13} s_{13} \sin\delta \sin\left(\frac{\Delta m^2_{32} t}{2E}\right) \\ 
        & \times e^{-\sigma \omega^4 (\Delta m^2_{32})^2 (\Delta m^2_{31} + \Delta m^2_{21})^2 t}.
    \end{aligned}
    \label{k}
\end{equation}
Equations (\ref{l}) and (\ref{k}) can be reformulated into a more general and compact expression
\begin{equation}
P_{\nu_A \nu_B} = \delta_{AB} + \sum_{j > k} \left[ C_{jk}(AB) + I_{jk}(AB) e^{-\Gamma_{jk} t} \right],
\label{proba}
\end{equation}
where the components are defined as follows: 
\begin{equation}
C_{jk}(AB) = -2 \operatorname{Re}(U_{B j} U_{A j} U_{A k} U_{B k}),
\end{equation}
and 
\begin{align}
I_{jk}(AB) &= 2 \operatorname{Re}(U_{B j} U_{A j} U_{A k} U_{B k}) \cos\left( \frac{\Delta m_{jk}^2}{2E} t \right) \nonumber \\
&\quad + 2 \operatorname{Im}(U_{B j} U_{A j} U_{A k} U_{B k}) \sin\left( \frac{\Delta m_{jk}^2}{2E} t \right).
\end{align}
Here, we can use \( t \sim L \) in Eqs. (\ref{l}) and (\ref{k}) while keeping \( c = 1 \). It is worth noting that the final expressions for the survival probability amplitude and the transition probability include an exponential damping factor, where \( L \) represents the neutrino oscillation path length. This is a common feature when the interaction of the neutrino subsystem with the environment, described by a dissipative term in the evolution of the reduced-density matrix, leads to damping effects in the oscillation probabilities. These effects are characterized by a factor \( e^{-\Gamma_{ij} L} \), where \( \Gamma_{ij} \) represents the damping strength. Consequently, the coherence length is given by \( l^{ij}_{\text{ch}} = \frac{1}{\Gamma_{ij}} \) \cite{set1}. Note
also, that for the inverted hierarchy case, the effect of decoherence on both survival and transition probabilities remains unchanged, as emphasized in \cite{gupdeco}, where the formulation is based on the Generalized Uncertainty Principle (GUP) approach \cite{gwt}. This observation also applies to our case, as the dissipator operator (\ref{myeq11}) remains invariant regardless of the mass hierarchy.

\section{Energy Dependence of Decoherence in \texorpdfstring{\(\kappa\)}{kappa}-Minkowski Spacetime }

We are now in a position to investigate the energy dependence of the decoherence parameter \(\Gamma_{ij} \sim \Gamma\) resulting
from the stochastic behavior associated with quantum
spacetime in the \(\kappa\)-Minkowski framework. Typically, decoherence models involving unknown environments suggest that interactions between the neutrino system and the quantum environment induce damping effects in oscillations. As discussed previously, recent stringent constraints on decoherence parameters with positive energy dependence (\(\Gamma \propto E^n\), where \(n > 0\)) have been applied to studies of atmospheric neutrinos observed at the IceCube Neutrino Observatory \cite{lisi3}. If the decoherence effects are indeed linked to quantum gravity, it is widely assumed that the exponent \(n\) should be positive \cite{sar}.

The assumption of a positive power-law dependence is predicted by quantum gravity models, such as those informed by effective field theories, string theory, and loop quantum gravity \cite{bd4,bd5,bc5,bd6}. These models are built on the premise that higher-energy (or shorter-wavelength) particles are more sensitive to spacetime fluctuations. The rationale is that Planck-scale fluctuations exert a stronger influence on shorter wavelength (i.e., higher-energy) particles compared to longer wavelength (i.e., lower-energy) ones. This assumption leads to a positive power law in decoherence, where the strength of decoherence increases with the particle energy.

Now, in our model, the decoherence parameter $\Gamma_{ij}(E)$ in Eq.~(\ref{proba})
is expressed in the usual power–law form
\begin{equation}
    \Gamma_{ij}(E)=\Gamma_{ij}(E_{0})\left(\frac{E}{E_{0}}\right)^{n},
    \label{sc}
\end{equation}
where $E_{0}=1\,\mathrm{GeV}$ is a commonly used pivot scale 
\cite{bd66,bd666,bd6666}.

Using the expressions derived from Eqs.~(\ref{l}) and (\ref{k}), the decoherence
rates for the three mass–eigenstate pairs are all of the same functional form
and differ only by the appropriate mass–squared combination.  
It is therefore convenient to write them in the compact, unified form
\begin{equation}
\label{Gamma_general}
    \Gamma_{ij}(E)
    =\frac{1}{16}\,\chi\, t_{P}\, a_{0}^{2}\,
      \big(\Delta m^{2}_{ij}\big)^{4}\, E^{-4},
\end{equation}
where 
\[
\Delta m^{2}_{21}=m_{2}^{2}-m_{1}^{2},\qquad
\Delta m^{2}_{31}=m_{3}^{2}-m_{1}^{2},\qquad
\Delta m^{2}_{32}=m_{3}^{2}-m_{2}^{2}.
\]

Comparing Eq.~\eqref{Gamma_general} with the general scaling form above Eq.~ (\ref{sc}), we
immediately identify the power–law index
\begin{equation}
    n = -4,
\end{equation}
Thus the decoherence rate decreases rapidly with increasing energy and becomes
most relevant in the low–energy regime.
%so that 
%\begin{equation}
   % \Gamma_{ij}(E)
  %  =\Gamma_{ij}(E_{0})\left(\frac{E}{E_{0}}\right)^{-4}.
%^\end{equation}
%Thus the decoherence rate decreases rapidly with increasing energy and becomes
%most relevant in the low–energy regime.}

%Now, in our model, the decoherence parameter $\Gamma_{ij}(E)$ in equation (\ref{proba}) is written in the standard scaling form
%\begin{equation}
    %\Gamma_{ij}(E) = \Gamma_{ij}(E_0)\left(\frac{E}{E_0}\right)^{n}, \quad i,j=1,2,3,
%\end{equation}
%where $E_0$ is a reference (pivot) energy scale, which we take to be $E_0 = 1$ GeV following common practice \cite{bd66,bd666,bd6666}. 
%From the explicit expressions obtained from equations (\ref{l}) and (\ref{k}), we have
%\begin{equation}
  %  \Gamma_{21}(E) = \frac{1}{16}\,\chi\,t_{P}\,a_{0}^{2}\,(\Delta m^{2}_{21})^{4}\,E^{-4}, 
   % \label{1}
%\end{equation}
%\begin{equation}
   % \Gamma_{31}(E) = \frac{1}{16}\,\chi\,t_{P}\,a_{0}^{2}\,(\Delta m^{2}_{31})^{4}\,E^{-4},
   % \label{2}
%\end{equation}
%\begin{equation}
   % \Gamma_{32}(E) = \frac{1}{16}\,\chi\,t_{P}\,a_{0}^{2}\,(\Delta m^{2}_{31} + \Delta m^{2}_{21})^{2}\,E^{-4}.
       % \label{3}
%\end{equation}
%Comparing with the general scaling form, it follows that
%\begin{equation}
   % n = -4,
%\end{equation}
%so that
%\begin{equation}
    %\Gamma_{ij}(E)=\Gamma_{ij}(E_0)\left(\frac{E}{E_0}\right)^{-4}.
%\end{equation}
%This makes clear that $\Gamma_{ij}(E)$ decreases with increasing neutrino energy and becomes most relevant in the low-energy regime.

It follows from the above expressions that the decoherence induced by quantum spacetime becomes more significant at lower neutrino energies, which contrasts with typical quantum gravity-induced decoherence models, where the effects tend to diminish at low energies. Interestingly, this result is similar to the dependence on the power law observed in neutrino decoherence due to wave packet separation in reactor experiments \cite{g1} and light-cone fluctuations \cite{g2}. Moreover, it has recently been suggested that scenarios involving extreme energy dependence (for example, \(n \leq -10\)) could potentially explain the Gallium anomaly \cite{bd}.

In order to estimate the coherence path length associated with \(\kappa\)-Minkowski spacetime fluctuations in natural units, we constrain the parameter \(\chi\), a phenomenological quantity, based on observable quantum spacetime effects. Spacetime coordinates are expressed through a perturbative expansion in terms of \(a_0\), assumed to be small. This perturbative expansion modifies the Heisenberg commutation relation (\ref{comm-NC}), with the requirement that these modifications remain negligible compared to the standard term \(i \hbar \eta_{\mu \nu}\). Thus, we impose the condition
\begin{equation}
|s \cdot a_0 \cdot p_0| < 1.
\end{equation}

%Now, in our model, the decoherence parameter \(\Gamma_{ij}(E)\) in equation (\ref{proba}) can be identified as \begin{equation} \Gamma_{ij}(E) = \Gamma_{ij}(E_0) \left(\frac{E}{E_0}\right)^n, \quad i,j=1,2,3. \end{equation} Here \(E_0\) is the pivot energy scale, which we set to \(E_0 = 1\) GeV, as commonly used in the literature \cite{bd66,bd666,bd6666}, and \(n\) is the power law index. Specifically, the explicit expression for the decoherence parameters can be derived from equations (\ref{l}) and (\ref{k}) as follows: \begin{equation} \Gamma_{21}(E) = \frac{1}{16} \chi t_{P} a_{0}^{2} (\Delta m^{2}_{21})^{4} E^{-4}, \label{1} \end{equation} \begin{equation} \Gamma_{31}(E) = \frac{1}{16} \chi t_{P} a_{0}^{2} (\Delta m^{2}_{31})^{4} E^{-4}, \label{2} \end{equation} \begin{equation} \Gamma_{32}(E) = \frac{1}{16} \chi t_{P} a_{0}^{2} (\Delta m^{2}_{31} + \Delta m^{2}_{21})^{2} E^{-4}. \label{3} \end{equation} From these expressions, it is clear that in our model, the power-law index is \(n = -4\).\\

Given that the decoherence parameters in Eq. (\ref{Gamma_general}) highlight the enhanced role of decoherence in the low-energy regime, we analyze the scenario using typical low-energy cosmic neutrinos characterized by \(p_0 = E \sim 10^{-6} \, \text{eV}\) \cite{bd1} and \(a_0 \approx 10^{-28} \, \text{eV}^{-1}\) \cite{d1}. This yields
$s < 10^{34}$. Now,
since \(\chi = (s + 1)^2\), we find
$\chi < (10^{34} + 1)^2 \approx 10^{68}$.

\begin{figure}[t!]

    \centering
    \includegraphics[scale=0.75]
    {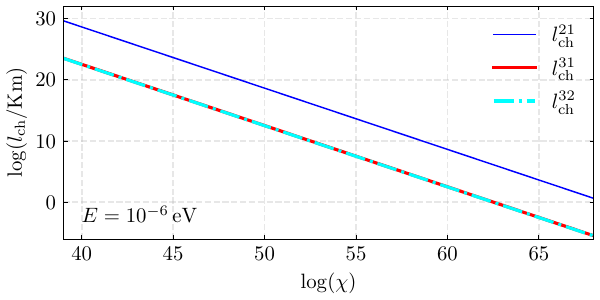}
    \caption{ Coherence length $l_{\mathrm{ch}}^{ij}$ as a function of the parameter $\chi$ at $E = 10^{-6}\,\mathrm{eV}$. 
The three curves correspond to the three independent neutrino mass-squared differences:
$l_{\mathrm{ch}}^{21}$ (blue), $l_{\mathrm{ch}}^{31}$ (red), and $l_{\mathrm{ch}}^{32}$ (green). 
Throughout the text, we use $l_{\mathrm{ch}}$ to denote the coherence length generically and 
$l_{\mathrm{ch}}^{ij}$ when referring to a specific mass-eigenstate pair $(i,j)$.}

    %Variation of the coherence path length with respect to \(\chi\) within the bounded region.}
   \label{fig}
\end{figure}

To estimate the parameter range of interest, we consider a representative coherence length $l_{\mathrm{ch}}$. Here we use $l_{\mathrm{ch}}$ (without the index $ij$) to denote an order-of-magnitude coherence scale, since all three $l_{\mathrm{ch}}^{ij}$ exhibit the same parametric dependence on $E$, $\chi$, $a_{0}$, and $\Delta m^{2}$ in the low-energy regime considered here. The comparison with the size of the observable Universe is used only as a conservative detectability reference scale and not as a fundamental theoretical bound on $l_{\mathrm{ch}}$. Since the neutrino is detected at $z=0$  (the present cosmological epoch), we compare $l_{\mathrm{ch}}(E)$ with the present-day particle horizon distance $D_H(z=0) \sim 10^{28}\,\mathrm{m}$ to avoid assuming coherent flavor evolution over distances beyond any causally connected region \cite{Planck:2018vyg}. This comparison therefore acts only as an order-of-magnitude consistency check.

More physically, one may instead require
\[
l_{\mathrm{ch}}(E) \lesssim l_{\mathrm{src}},
\]
where $l_{\mathrm{src}}$ denotes a characteristic source–detector separation, such as a Galactic ($\sim 50$ kpc) or extragalactic (Mpc–Gpc) baseline \cite{Planck:2018vyg}. This avoids any reliance on cosmological horizon scales and leads to the same qualitative constraints in the low-energy regime considered here. Note also that while the cosmological horizon evolves with time, $l_{\mathrm{ch}}(E)$ is defined at the detection energy $E(z=0)$ and is therefore constant for the observed neutrino.

Using the representative expression for the coherence length derived from the inverse of $\Gamma$ in Eq. (\ref{Gamma_general}),
\begin{equation}
l^{ij}_{\mathrm{ch}} \sim \frac{E^{4}}{\chi\,a_{0}^{2}\,t_{p}\,(\Delta m^{2}_{ij})^{4}},
\end{equation}
and taking $E \approx 10^{-6}\,\mathrm{eV}$, $a_{0}^{2} = 10^{-56}\,\mathrm{eV}^{-2}$, $t_{p} = 10^{-28}\,\mathrm{eV}^{-1}$, and $\Delta m^{2} = 10^{-3}\,\mathrm{eV}^{2}$ \cite{bd1}, we obtain the lower bound $\chi > 10^{39}$. Combining this with the upper bound derived from the perturbative condition, we find $
10^{39} < \chi < 10^{68}$.
For a slightly higher energy ($E \sim 10^{-2}\,\mathrm{eV}$), the bounds shift to $10^{55} < \chi < 10^{60}$.

At high energies, the lower bound inferred from $l_{\mathrm{ch}}$ can exceed the perturbative upper bound, which suggests a limitation in the applicability of the present model in the high-energy regime.

%To establish a lower bound, we require that the coherence length \(l_{\text{coh}}\) remain smaller than the size of the observable universe, i.e. \(L_{\text{coh}} < 10^{28} \, \text{m}\), or equivalently \(l_{\text{coh}} < 10^{33} \, \text{eV}^{-1}\) \cite{bd2}. The coherence length is typically determined from the inverse of the decay factor \(\Gamma\) in equations (\ref{1}), (\ref{2}), and (\ref{3}), as follows:
%\begin{equation}
%l_{\text{coh}} \sim \frac{E^4}{\chi \cdot a_0^2 \cdot t_p \cdot (\Delta m^2)^4}
%\end{equation}
%Using \(E \approx 10^{-6} \, \text{eV}\), \(a_0^2 = 10^{-56} \, \text{eV}^{-2}\), \(t_p = 10^{-28} \, \text{eV}^{-1}\), and \(\Delta m^2 = 10^{-3} \, \text{eV}^2\) \cite{bd1}, we deduce $\chi > 10^{39}$.
%Combining these bounds, the  range for \(\chi\) is 
%$10^{39} < \chi < 10^{68}$.
%Similarly, for a slightly higher neutrino energy (\(E \sim 10^{-2} \, \text{eV}\)), the bounds for \(\chi\) adjust to
%$10^{55} < \chi < 10^{60}$.% Interestingly, in the high-energy regime, the %lower bound derived from $L_{\text{coh}} < 10^{33} \, \text{eV}^{-1}$ exceeds the upper bound obtained by using the perturbative approximation. This suggests a potential limitation in the model's applicability for high-energy neutrinos.

\begin{figure}[t!]
    \centering
    \includegraphics[scale=0.75]
    {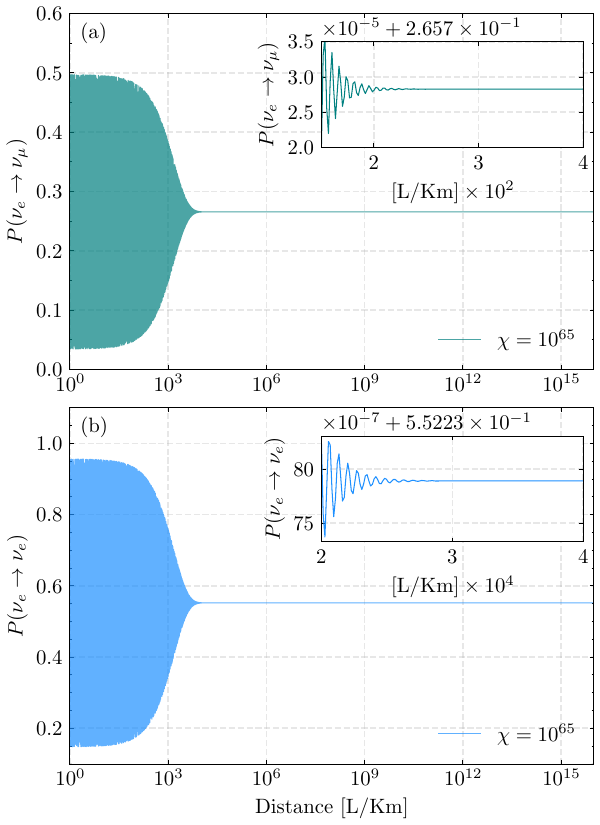}
    \caption{Decoherence effect on standard survival and transition probabilities of C$\nu$B neutrinos for \(\nu_{e} \rightarrow \nu_{e}\) (depicted in sky blue) and \(\nu_{e} \rightarrow \nu_{\mu}\) (depicted in green), are plotted as a function of path length \( L \) (in km) at a fixed neutrino energy \( E = 10^{-6} \, \text{eV}\).}
   \label{fig:cnB-survival-probability}
\end{figure}

\begin{figure}[t!]
    \centering
    \includegraphics[scale=0.75]{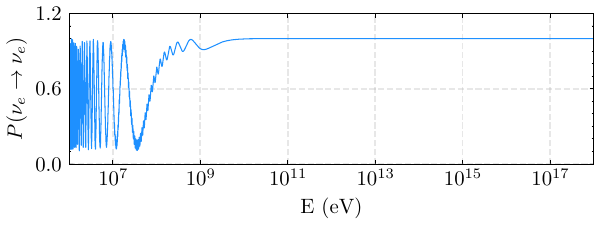}
    \caption{Standard C$\nu$B neutrino survival probability for \(\nu_{e} \rightarrow \nu_{e}\) (\ref{l}) plotted as a function of Energy \( E \) (measured in eV) for a fixed  \( \chi = 10^{62} \) and path length \( L \sim 10^{14} \, \text{km} \).}
    \label{fig:cnB-survival-probabilit}
\end{figure}

\begin{figure}[t!]
    \centering
    \includegraphics[scale=0.75]{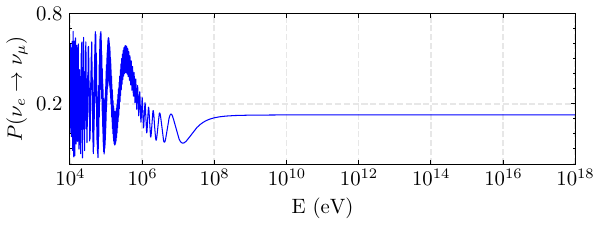}
    \caption{Standard C$\nu$B neutrino transition probability for \(\nu_{e} \rightarrow \nu_{\mu}\) (\ref{l}) plotted as a function of Energy \( E \) (measured in eV) for a fixed  \( \chi = 10^{62} \) and path length \( L \sim 10^{14} \, \text{km} \).}
    \label{fig:cnB-transition-probabilit}
\end{figure}

In Fig.~\ref{fig} we plot the representative coherence length $L_{\mathrm{coh}}$ as a function of $\chi$, confirming that larger values of $\chi$ enhance the decoherence strength. The dependence of $\chi$ on neutrino energy follows the scaling $\Gamma_{ij}(E)\propto E^{-4}$ discussed above, implying stronger decoherence at lower energies. The corresponding effects on the survival and transition probabilities at fixed energy $E$ are shown in Fig.~\ref{fig:cnB-survival-probability}, while the dependence on $E$ is illustrated in Figs.~\ref{fig:cnB-survival-probabilit} and \ref{fig:cnB-transition-probabilit}. These results demonstrate a characteristic feature of our framework: decoherence is strongly suppressed at high energies and becomes increasingly significant in the infrared. At sufficiently high energies, the oscillation curves converge to the standard unitary predictions, consistent with IceCube observations \cite{lisi1,lisi2,lisi3}, which place strong constraints on models predicting decoherence that grows with increasing energy.

The reference to relic--neutrino energies in this work is not intended to suggest that
the cosmic neutrino background (C$\nu$B) exhibits observable flavour oscillations.
Relic neutrinos today form an incoherent mixture of mass eigenstates due to early-universe
wave--packet separation and other decohering effects, and therefore do not oscillate in the
standard sense.  
Our use of the C$\nu$B energy scale serves only to illustrate the strong infrared behaviour implied
by the $E^{-4}$ dependence of the decoherence parameter.

Importantly, the infrared enhancement of quantum--spacetime--induced decoherence does
\emph{not} rely on flavour oscillations being observable.
Even for an incoherent ensemble, decoherence modifies the evolution of the reduced density matrix,
and such modifications can leave indirect signatures in detection processes that depend on the
flavour projection at the detector.  
In particular, in neutrino capture on tritium---as relevant for PTOLEMY and related proposals
\cite{pot}---the capture rate depends on $\mathrm{Tr}(P_e \rho_f)$,
i.e.\ on the flavour-projected density matrix rather than on the presence of flavour oscillations
themselves.  
This distinction is well established in the C$\nu$B capture literature and implies that
decoherence effects may still influence the capture signal even when oscillations are unobservable.  
A detailed derivation of how quantum--spacetime--induced decoherence enters the C$\nu$B capture
cross section lies beyond the scope of the present work.
Such an analysis is currently underway, building on the framework of 
Ref.~\cite{Capolupo:2023xek}, and will be presented elsewhere~\cite{Nandi:InPrep}.

%\textcolor{red}{
%It is important to emphasize that infrared enhancement of quantum--spacetime--induced decoherence does
%\emph{not} rely on flavour oscillations being observable.  
%Even for an incoherent ensemble, decoherence modifies the evolution of the reduced density matrix, and
%such modifications can leave indirect signatures in detection processes that depend on the flavour
%projection at the detector.  
%In particular, in neutrino capture on tritium---as relevant for PTOLEMY and related proposals\cite{pot}---the
%capture rate is proportional to $P(\nu_A\rightarrow \nu_B;t)$, and therefore depends on the flavour-projected
%density matrix rather than on flavour oscillations themselves.  
%This distinction is well known in the C$\nu$B capture literature and implies that decoherence effects
%may still influence the capture rate even when oscillations are unobservable.  
%A detailed derivation of how quantum--spacetime--induced decoherence enters the C$\nu$B capture
%cross section is beyond the scope of the present paper.  
%Such an analysis is currently underway, building upon the framework of
%Ref.~\cite{Capolupo:2023xek}, and will be presented %elsewhere~\cite{Nandi:InPrep}. 
%}

Therefore, the relic–neutrino energy scale is used here only as a benchmark to highlight
the strong low–energy enhancement of decoherence predicted by the model.
The phenomenologically relevant regime for testing this effect is any setting in which
neutrinos are produced coherently—such as laboratory, reactor, beta--decay, or solar
neutrinos—so that flavour evolution and its decoherence can be meaningfully probed.

\section{Conclusions}

The present study examines decoherence effects induced by fluctuations in the stochastic, non-commutative \(\kappa\)-Minkowski spacetime at the Planck scale. Our results demonstrate that the stochastic nature of quantum spacetime, with the deformation parameter \(a_{0}\) introducing an inherent fuzziness at the Planck scale, can itself serve as a source of decoherence, establishing a clear connection between the flat limit of quantum gravity and neutrino oscillations. This intrinsic fuzziness contributes an additional term to the Lindblad master equation, describing the interaction between neutrinos and quantum spacetime.
Specifically, we analyze how the fluctuations of the \(\kappa\) Minkowski spacetime affect the decoherence parameter. 

While $\kappa$-Minkowski non-commutativity replaces the smooth spacetime manifold with a Lorentz-covariant Lie algebraic structure at the Planck scale, it also modifies the conventional Heisenberg phase space algebra, resulting in a non-flat momentum space \cite{dd,dd1}. Notably, this non-flat momentum space allows Planck-scale effects to manifest over extended distances, even for low-energy particles. The decoherence arising from these fluctuations is interpreted as an averaged effect, emerging from the cumulative interaction with spacetime variations. For low-energy particles, these stochastic fluctuations induce small, localized disturbances in the quantum state. Although each individual interaction is weak, their effects accumulate over long distances, leading to significant decoherence \cite{d2,bd3}.
      
The resulting decoherence factor exhibits a distinctive energy dependence, scaling as \(\Gamma \propto E^{-4}\). This feature stems from a nuanced interplay between the interaction strength and the propagation dynamics. Although high-energy neutrinos interact more strongly with space-time fluctuations because of their shorter wavelengths, their rapid propagation limits the cumulative decoherence effects over time or distance. The cumulative nature of spacetime interactions in the \(\kappa\)-Minkowski spacetime offers a compelling explanation for the predicted inverse energy scaling of decoherence in our model.  Low-energy neutrinos, characterized by longer wavelengths and slower propagation speeds, are more susceptible to cumulative spacetime fluctuations, leading to amplified decoherence.

The negative power-law dependence of the decoherence parameter distinguishes decoherence effects arising from unknown environmental factors from those predicted by quantum-gravity models. Specifically, our model identifies an exponent \( n = -4 \) as a characteristic signature of \(\kappa\)-Minkowski–type noncommutative spacetime, where stochastic fluctuations at the Planck scale drive decoherence. This scaling implies that quantum-spacetime effects are more significant for low-energy neutrinos than for their high-energy counterparts. In this sense, the relic-neutrino energy scale serves only as an infrared benchmark for illustrating the strong \( \Gamma \propto E^{-4} \) enhancement, rather than suggesting that relic neutrinos themselves undergo observable flavour oscillations today. A notable illustrative example is the cosmic neutrino background (C\(\nu\)B), produced under ultra-relativistic early-universe conditions; as the Universe expanded and neutrinos cooled, they entered the low-energy regime where the inverse-power behaviour amplifies the cumulative effect of spacetime fluctuations.

Our results are fully consistent with IceCube’s observations of atmospheric neutrinos, which exhibit no appreciable loss of coherence over TeV energies and long baselines \cite{lisi1,lisi2,lisi3}. This supports the expectation that high-energy neutrinos are essentially insensitive to the \(\kappa\)-Minkowski–induced decoherence considered here. Recent IceCube analyses \cite{lisi3} place strong bounds on quantum-gravity–induced decoherence at high energies, and these bounds are naturally satisfied in our framework due to the \( \Gamma \propto E^{-4} \) suppression. In contrast, the strongest effects are predicted at very low energies, suggesting that the meV-scale regime relevant for C\(\nu\)B \emph{capture} experiments may provide an experimentally accessible window into Planck-scale physics. While relic neutrinos do not oscillate today due to early-universe decoherence, the capture process depends on the flavour-projected density matrix and may therefore retain indirect signatures of quantum-spacetime fluctuations.

An alternative approach related to decoherence mechanisms in \(\kappa\)-Minkowski spacetime has been explored in \cite{b17}, where the deformation is incorporated at the level of the translation generators, leading to a deformed coproduct in the Hopf algebra structure \cite{ori5}. In this formulation, the modified symmetry algebra results in a non-unitary evolution of the density matrix, without requiring any additional stochastic noise terms. Decoherence, in this case, is an intrinsic feature of the algebraic structure of quantum spacetime.  

In contrast, we adopt a canonical phase-space realization, a common strategy in quantum spacetime models \cite{RB}, particularly for the \(\kappa\)-deformed phase-space algebra (\ref{kappa}, \ref{comm-NC}). This approach expresses noncommutative phase-space variables in terms of laboratory-frame canonical variables \cite{b18}, which are the directly measurable quantities used by experimentalists. By treating noncommutative effects as effective corrections, this framework enables meaningful comparisons with real-world data, ensuring that deviations from standard neutrino oscillations are expressed in experimentally interpretable terms. Unlike the deformed coproduct approach, where non-unitarity is inherent from the Hopf-algebraic perspective, our framework preserves unitary evolution by working with canonical variables, with decoherence arising only if quantum spacetime fluctuations introduce stochasticity.  

Both approaches provide valuable insights into how quantum spacetime may influence the decoherence mechanism, but their distinction remains an empirical question that must be addressed through experimental constraints on neutrino coherence across different energy scales. Since neutrino oscillation experiments are fundamentally formulated in terms of canonical commutative phase-space variables, we argue that, unlike the predictions in \cite{b17}, which suggest that decoherence increases with energy, our framework naturally explains IceCube's null results by predicting a strong suppression of decoherence at high energies. This indicates that next-generation low-energy neutrino observatories, particularly those designed to probe relic neutrinos, could provide the most promising avenue for testing quantum gravity-induced decoherence.

We conclude with some remarks on the prospects for further study. Our approach to investigating decoherence from quantum spacetime can be extended to more general formulations of quantum spacetime beyond the \(\kappa\)-Minkowski framework \cite{b+}. A broader geometric setup has recently been developed \cite{ori6}, which could provide insights into potential modifications of the power-law behavior of the decoherence parameter.

Current and next-generation high-energy neutrino telescopes, such as IceCube \cite{lisi3}, IceCube-Gen2 \cite{bd8}, and KM3NeT \cite{bd9}, have placed strong constraints on quantum gravity-induced decoherence at TeV energies. However, since our model predicts that decoherence effects are strongest at low energies, these high-energy telescopes are not ideal for directly testing this prediction. Current and next-generation high-energy neutrino telescopes, such as IceCube \cite{lisi3}, IceCube-Gen2 \cite{bd8}, and KM3NeT \cite{bd9}, have placed strong constraints on quantum gravity-induced decoherence at TeV energies. However, since our model predicts that decoherence effects are strongest at low energies, these high-energy telescopes are not ideal for directly testing this prediction. Instead, future low-energy neutrino observatories—whether based on beta-decay, reactor, solar, or C\(\nu\)B-sensitive technologies—may offer the most promising avenue for probing Planck-scale–induced decoherence.
% Instead, future low-energy neutrino observatories—particularly those pursuing cosmic-neutrino-background (C\(\nu\)B) \emph{capture} measurements—may offer an indirect window into Planck-scale effects on neutrino coherence.

One promising experiment in this direction is PTOLEMY \cite{pot}, which aims to detect relic neutrinos from the early universe in the meV energy range and could serve as a crucial test of quantum gravity-induced decoherence. Detecting signatures of quantum spacetime remains a formidable challenge, yet this work lays the foundation for future explorations in this direction. The interplay between quantum and gravitational effects is essential, as it may play a pivotal role in the pursuit of a theory of quantum gravity.

\section*{Acknowledgements}
PN acknowledges the support of the Rector’s Postdoctoral Fellowship Program (RPFP) at Stellenbosch University and is grateful to Prof. Frederik G. Scholtz for insightful discussions. PN and TB also appreciate the valuable correspondence with Ms. Nandita Debnath, Mr. Subhajit Kala, and Mr. Indra Kumar Banerjee. PN would further like to thank Prof. Antonio Capolupo (University of Salerno and INFN, Salerno) for insightful correspondence. Finally, the authors thank the anonymous reviewers for their constructive comments, which helped improve the clarity of this work. %\textcolor{red}{GP and FP funding.}. 

%\begin{thebibliography}{0}

\end{document}